\documentclass[a4paper]{jpconf}
\usepackage{graphicx}
\begin{document}
\title{Galactic sources of  high energy neutrinos}

\author{Felix  Aharonian}

\address{Dublin Institute for Advanced Studies, 5 Merrion Square,
Dublin 2,  Ireland \&  \\
Max Planck Institut f\"ur Kernphysik, Saupfercheckweg 1, 69117 Heidelberg, Germany}

\ead{felix.aharonian@mpi-hd.mpg.de}

\begin{abstract}
The undisputed galactic origin of  cosmic rays at energies below
the so-called knee around $10^{15} \rm eV$ implies an existence of  
a nonthemal  population of  galactic objects which  
effectively accelerate 
protons and nuclei to  TeV-PeV energies.  
The distinct signatures of these cosmic accelerators
are high energy  neutrinos and $\gamma$-rays 
produced through  hadronic interactions.  
While $\gamma$-rays can be produced also  
by directly accelerated electrons, high energy neutrinos 
provide the most straightforward and 
unambiguous information  about the nucleonic component of accelerated 
particles.  The planned  $\rm km^3$-volume class high 
energy neutrino detectors are expected to be  sensitive enough to 
provide  the first astrophysically  meaningful probes of  
potential VHE neutrino sources.
This  optimistic prediction  is based on the recent 
discovery of high energy  $\gamma$-ray sources 
with hard energy  spectra  extending to 10 TeV and beyond.   
Amongst the best-bet  candidates are two young shell-type 
supernova remnants -- RXJ~1713.7-4946 and RXJ~0852.0-4622, 
and perhaps also  two prominent plerions  - the Crab Nebula and Vela X. 
Because of  strong absorption of TeV $\gamma$-rays, one may expect  
detectable neutrino fluxes also from  (somewhat  fainter) 
compact TeV   $\gamma$-ray emitters 
like  the binary systems LS~5039 and LS~I+61~303, and, hopefully, 
from  hypothetical "hidden" or "orphan" neutrino sources. 
\end{abstract}

\section{Introduction}

Very High Energy   ($E \geq 0.1$~TeV; VHE) neutrinos are unique messengers of nonthermal phenomena  in the Universe related to the hadronic interactions of protons and nuclei  in cosmic TeVatrons and 
PeVatrons -  Nature's masterly designed  machines accelerating particles to 
TeV and PeV energies.  In this regard VHE neutrinos are complementary to $\gamma$-rays which are produced both in electromagnetic and hadronic 
interactions.  On the other hand, unlike $\gamma$-rays, 
neutrinos are not fragile;
they interact only weakly with the ambient medium - gas, 
radiation and magnetic fields, 
and thus carry information about high energy processes
occurring  in  ''hidden''  
regions where the particle accelerators could be located. This concerns, first of all, the regions associated with compact objects - 
black holes, pulsars, 
the initial epochs of supernovae explosions, \textit{etc}.  The  penetrating potential
of neutrinos is important not only for extremely dense environments in which $\gamma$-rays are dramatically absorbed,
but also moderately opaque sources from which we can see $\gamma$-rays, but 
after significant distortion due to internal and external absorption. 

Ironically, this nice  (from an astrophysical point of view)  feature
of neutrinos  makes, at the same time,  their detection extremely difficult. 
This explains  why 
over several decades high energy neutrino astronomy has remained
essentially  a theoretical discipline with many  exciting ideas and 
predictions but without  the  detection of a single VHE neutrinos source. 
However, it is  expected that, with arrival of 
the km$^3$-volume class scale  detectors like IceCube and KM3NeT 
(see e.g. refs. \cite{Science_Halzen,Katz}),  
the status of the field  will  be changed dramatically. Generally, 
predictions of VHE neutrino fluxes from astrophysical objects 
rely on assumptions with not well-constrained model parameters and, 
therefore,  often 
contain large  (orders of magnitude!) uncertainties.  This leaves a 
significant  freedom in speculations  on the   
"best-bet neutrino sources",
and consequently  allows  a broad spectrum of opinions 
concerning the prospects for detecting  the first 
astrophysical neutrinos -  from very enthusiastic  statements to 
rather careful predictions prevailed by 
a healthy scepticism (see e.g. ref. \cite{Lipari}). 
 
Presently extragalactic objects like Active Galactic Nuclei (AGN) and 
sources of Gamma Ray Bursts (GRBs) are believed to be the most likely 
objects  to be detected as neutrino sources, and therefore the driving 
motivation  of experimental VHE neutrino astronomy
(see e.g. ref. \cite{Waxman}). The current models of AGN and GRBs  indeed
contain many attractive components (concerning the conditions of particle 
acceleration and their interactions; see e.g. ref. \cite{Rachen98}) 
that makes these objects 
\textit{potentially detectable}  sources of VHE neutrinos 
(see e.g. refs. \cite{Meszaros,DerAt}). 
On the other hand,
the poor understanding of certain aspects of the  physics of AGN and especially GRBs, 
as well as the lack of  constraints on neutrino productions rates from $\gamma$-ray
observations  (because of intrinsic and intergalactic absorption of 
VHE  $\gamma$-rays),  formally  allow  calculations
in extreme model-parameter segments which  
often lead  to rather high (in the case of Blazars - overoptimistic)  
neutrino flux predictions.

In contrast to extragalactic objects, 
the models of potential galactic neutrino sources, in particular
the shell type Supernova Remnants (SNRs), Pulsar Wind Nebulae (PWNe),
Star Formations Regions and the 
dense molecular clouds related to them,  
are robustly  constrained 
by $\gamma$-ray observations of the galactic disk 
in very-high  ($\geq 1$ TeV) \cite{HESS_survey,Milagro} 
and ultra-high energy ($\geq 100$ TeV) \cite{CASA-MIA} domains. 
Typically, the expected fluxes  from these objects are below the detection 
threshold of the planned neutrino detectors. However,  
the recent HESS discoveries  of  several TeV $\gamma$-ray sources 
at the flux level of "1 Crab", which can be interpreted within the
hadronic models of gamma-ray emission, 
sustain a hope that first TeV galactic sources  
will be detected in foreseeable future by km$^3$-volume class 
instruments like IceCube  and Km3NeT.

\section{On the detectability of galactic VHE neutrino sources}

The recent performace studies 
of the km$^3$-volume scale detectors  
show that the detection of persistent point-like  
neutrino sources (for a typical 
angular resolution of VHE neutrino detectors the "point-like" source 
implies an object of angular size $\leq 1^\circ$)
for a realistic  exposure time 
(typically, a  few years  continuous observations) 
is limited by a flux 
$F (\ge  1 \rm TeV) \approx  10^{-11} \  \rm \nu/cm^2 s$ (see e.g. refs. \cite{Vissani,Beacom,Distefano,Kappes}). 
The corresponding energy flux is  
$f_{\rm E} \approx  10^{-10} \ \rm erg/cm^2 s$ 
or somewhat less, depending on the  
spectrum in the 
most relevant energy interval between 1 and 100 TeV.
This exceeds, by two orders of magnitude, the minimum 
$\gamma$-ray flux detectable in the same energy band.
On the other hand, the sensitivity of the 
km$^3$-scale  detectors is comparable or better than
the minimum  detectable  energy flux achieved by  
the Compton Gamma Ray Observatory detectors
(COMPTEL, EGRET) in the MeV/GeV $\gamma$-ray band. 
For an  isotropic VHE source located at a 
distance $d$,  
the luminosity of TeV neutrinos can be
probed at the level  
\begin{equation}
L_\nu \simeq 10^{34} (d/1 \rm \ kpc)^2 \ \rm erg/s  \ .
\end{equation}
At first glance,  this is a quite modest luminosity, at least  for a powerful 
hadronic source located in a dense environment.  Indeed, for production 
of TeV neutrinos in \textit{p-p }interactions with ambient 
gas of density  $n_0=n/1  \rm cm^{-3}$, the required total 
energy in multi-TeV protons  is estimated 
$W_p \simeq  t_{\rm pp}  c^{-1}_{\rm p \rightarrow \nu} 
L_\nu \simeq 10^{49}  n^{-1}_0  d_{\rm kpc}^2  \ \rm erg$, where 
$t_{\rm pp} \approx 10^{15} n_0^{-1} \rm \ s$ is the 
radiative cooling time of protons due to inelastic \textit{p-p} 
interactions, and $c_{\rm p \rightarrow \nu} \approx  0.1$ is 
the fraction of average energy of a proton transferred to 
neutrinos.  One may conclude that even in a 
relatively low density environment ($\rm n_0 \sim 1$)   
the required total energy can be readily produced 
in young SNRs through diffusive shock acceleration 
(see e.g. ref. \cite{Malkov}) or by a powerful pulsar 
assuming that a major  fraction of the spin-down 
luminosity of the pulsar is converted  to an 
ion-dominated wind  (see e.g. ref. \cite{Arons}). However,  this  
kind  of estimates can be misleading since 
they are based on a  silent assumption  that all particles accelerated 
during the life-time of the source  are effectively confined in a 
relatively compact region inside or nearby  the accelerator. 
In fact, production of TeV neutrinos requires
protons with energies well 
beyond 10 TeV  the escape of which from the source  
is difficult to prevent. This, of course,  would lead to 
a significant reduction of the neutrino production 
efficiency which in the case of \textit{p-p} interactions 
can be expressed in the following form: 
$\eta =L_\nu / \dot{W}_{\rm p}=  
{\rm min} [1,  t_{\rm esc}/t_{\rm pp}] \times c_{\rm p \rightarrow \nu}$, 
where $t_{\rm esc}$ is the escape time of nonthermal particles. If the escape proceeds in the 
diffusion regime,  then $t_{\rm esc}=R^2/2 D(E)$. It is convenient  
to write the diffusion coefficient $D(E)$ in the following  form 
$D(E)= \xi r_{\rm L} c/3 = 3.3 
\times 10^{22} \xi  E_{\rm TeV} B_{\rm mG}^{-1} \ \rm cm^2/s$, where 
$E_{\rm TeV}=E/1  \rm TeV$ is the proton energy normalized to 1 TeV, 
and  $B_{\rm mG}=B/10^{-3} \  \rm G$ is the magnetic field in units 
of mG.  Generally the parameter $\xi \geq 1 $ is a function of energy.  
In the most effective confinement regime corresponding to the 
Bohm diffusion,  $\xi=1$. Thus,   
in a source of a linear  size  $R_{\rm pc}=R/1 \ \rm pc$,  
the production efficiency  of TeV neutrinos, assuming  that a
hard energy spectrum of protons  extends effectively  to 100 TeV, is
\begin{equation}
\eta \approx 10^{-2} c_{p \rightarrow \nu} \xi^{-1} R_{\rm pc}^2 n_0 B_{\rm mG} \ .
\end{equation}
The maximum possible efficiency of a TeV neutrino source, 
$\eta \rightarrow c_{p \rightarrow \nu} \approx  0.1$  can in principle 
be achieved if 
$R_{\rm pc}^2 n_0 B_{\rm mG} \geq 10^2 \xi$. 
Such a condition can be best fulfilled in 
objects like giant molecular clouds,
with a size $R \sim 10$ pc, mass $10^5 M_\odot$, and magnetic field 
$B \geq  0.1$~mG, provided that the propagation  of multi-TeV 
protons proceeds close the Bohm diffusion 
regime.  However, Bohm diffusion  
hardly can be realized in molecular clouds, thus   
$\eta \ll  c_{p \rightarrow \nu}$, typically $\eta \leq 10^{-3}$,  
which implies that the acceleration power should exceed 
$\dot{W}_{\rm p} = \eta L_\nu \geq 10^{37}  d_{\rm kpc}^2 \ \rm erg/s$.
This significantly reduces the  
number of potentially detectable neutrino 
sources to the  most powerful  representatives of 
nonthermal source 
populations in our Galaxy.  
In addition to young SNRs and PWNe, possible emitters of TeV neutrinos 
are compact  binary  systems in which the compact object 
(a black hole or a pulsar) plays the role of 
particle accelerator, and the dense gas regions, 
e.g. the atmosphere of the companion 
star~\cite{Berezinsky,Vestrand,Hillas,Gaisser}
or the  accretion plasma around the 
compact object~\cite{Luis}, play the  role of the target. 

Moreover, in binary systems containing a luminous 
optical star  and a compact object, the photomeson interactions 
could provide an additional  channel for neutrino production, 
provided that protons are accelerated to energies  exceeding 
the interaction threshold, 
$E_{\rm th} \approx (200 \rm MeV/3kT) m_p c^2  \approx  10^4$~TeV 
in the case of interactions with the starlight, and three orders 
of magnitute less for interactions with photons of the accretion disk.  
For a photon field with a thermal (Planckian)  distribution, the   interaction and escape times of protons are
$t_{\rm esc} \approx 10^{10} \xi^{-1} R_{\rm pc}^2 B_{\rm mG} (kT/3 \rm eV)  \ \rm s$, and 
$t_{\rm p \gamma} \approx 10^{18} L_{37} R_{\rm pc}^2 (kT/3 \rm eV)  \ \rm s$, respectively.
It is remarkable  that both timescales are proportional, although for completely different reasons (!),   to the product  $R^2 kT$, 
therefore the neutrino production efficiency appears independent, 
for a given luminosity of thermal radiation,  of both the  
source size and the  
temperature of radiation, but strongly depends on the magnetic field 
\begin{equation}
\eta \approx  10^{-9} \xi^{-1} L_{37} B_{\rm mG} \ . 
\end{equation}
Thus we arrive at the conclusion that the neutrino production 
via photomeson interactions can be 
effective  only in compact objects  with strong  magnetic fields 
($B \geq 10$ kG) and high turbulence  (for confinement   
of protons  with $\xi \sim 1$).  The  immediate proximity of 
the luminous star (i.e. its photosphere) 
or the compact object, e.g   the accretion disk or the base of the jet~\cite{Levinson,Carla,Mitya},  
can be sites where the neutrino production 
proceeds with a reasonably high efficiency. 

\section{Detectability of neutrino sources in the 
context of multiwavelength observations} 
The \textit{high efficiency} of neutrino production is a 
key condition for the detectability of 
potential VHE neutrino sources  
given the \textit{limited budget} of available energy and the \textit{limitted sensitivity} of detectors. 
The above qualitative estimates show that this condition 
can be achieved only with  certain 
combinations of a few key model parameters.  Therefore, a careful inspection 
of these conditions  before  any  detailed numerical calculations   
is highly advisable. 

Independent constraints  on the detectability of  a  
VHE neutrino source  can be obtained also  
from analysis of multi-wavelength observations, in particular  
in the most relevant VHE $\gamma$-ray band. Indeed, 
TeV $\gamma$-ray fluxes  can be safely  used as upper limits for 
neutrino fluxes, provided, of course, that the internal 
and external absorption of $\gamma$-rays is negligible.  
Generally, the production of VHE neutrinos is accompanied by 
production  of $\gamma$-rays, but not \textit{vice versa};
$\gamma$-rays are 
copiously produced also in electromagnetic interaction both by electrons 
(through bremsstrahlung, inverse Compton scattering) and protons 
(e.g. through the synchrotron and curvature radiations).  Moreover, since 
the main  channels of the  neutrino and "hadronic"  gamma-ray production 
are decays  of  charged $\pi^\pm$ and neutral $\pi^0$ mesons,  with decay time of 
charged pions significantly longer than the decay time of neutral pions, 
at certain conditions the production of neutrinos can be suppressed 
compared to $\gamma$-ray production. 
Indeed,  in the lab frame, the decay time of
charged pions responsible for TeV neutrino production is
$t_{\pi^\pm}= (E_\pi/ m_\pi c^2)\,\tau_{\pi^\pm}
\approx 2.5 \times 10^{-3} (E_\pi/10 \ \rm TeV)$~s.
On the other hand, the cooling time  of pions 
due to inelastic $\pi p$ and $\pi \gamma$
interactions depends on densities of the ambient gas $n_p$
and X-ray photons $n_{\rm x}$:
$t_{\pi p} \sim 10^{14} (n_{\rm p}/1 {\rm cm^{-3}})^{-1}$ s,
and
$t_{\rm \pi \gamma} \sim 10^{18} {(\rm n_{\rm x} /1 
{\rm cm^{-3}}})^{-1}$~s, respectively. 
Thus, charged pions would decay to $\mu$ and $\nu_\mu$  
before interacting with the ambient photons and protons if 
$n_{\rm x} \leq  10^{21} \ \rm cm^{-3}$ and
$n_{\rm p} \leq  10^{17} \ \rm cm^{-3}$.
Finally, the production of neutrinos from the subsequent muon decay
would proceed with high probability as long as the magnetic
field does not exceed $B \approx 10^6$~G. This follows
directly from the comparison of the decay time of muons,
$t_\mu= (E_\mu/m_\mu c^2)\tau_{\mu} \simeq
0.2(E_\mu/10 \rm TeV)$~s, with their
synchrotron cooling time, 
$t_{\rm sy} \approx 0.07
\,(B/10^6 \rm G)^{-2} (E_\mu/10~{\rm TeV})^{-1}$~s.

Thus, the  ratio of $\gamma$-ray and neutrino production rates
in a VHE source  is expected to be of order of 1 or more,  
$\gamma/\nu \geq 1$.  The absorption of $\gamma$-rays 
can, of course,  significantly change the initial 
$\gamma/\nu$ ratio. While for galactic sources the external 
$\gamma$-ray absorption (due to interactions with the interstellar 
IR photon fields) is not dramatic up to 
several tens of TeV \cite{Mosk}, in compact galactic 
objects like X-ray binaries the internal absorption can 
be huge. However, the absorption 
of gamma-rays does not imply  that the information is lost; the 
secondary electrons initiate  a cascade in the same 
radiation field and/or cool via
synchrotron radiation. In the first case
(which happens 
when the radiation density $w_{\rm r}$ exceeds the  
energy density of the magnetic field  
$B^2/8 \pi$),  the initial $\gamma$-rays are 
gradually reprocessed, down to energies at which the 
source becomes transparent. In binary systems containing 
luminous optical stars (like LS~5039) the main energy 
is released at GeV energies \cite{Mitya,Bednarek1}, 
while in the case of particle acceleration in the accretion disk
around a black hole (like Cygnus X-1), the initial energy is 
released mainly at MeV energies \cite{Valera}. The cascade 
development in the photon field can be suppressed at the presence of 
a strong magnetic field, $B \geq \sqrt{8 \pi w_{\rm r}}$ 
(typically 10~G or more). Interestingly, in  this case 
the synchrotron radiation of pair-produced electrons 
extends well into the MeV gamma-ray energy band, 
and therefore its  detection is not prevented 
("screened")  by the optical and X-ray backgrounds 
related to the optical star and the compact object. Thus,
independent of the cooling regime
(via synchrotron radiation and/or the Klein-Nishina cascades), 
the pair-produced electrons leave a 
detectable imprint 
in the form of MeV to GeV gamma-ray photons.  
So far the MeV and 
GeV $\gamma$-ray sky has been explored at the  depth corresponding 
to the energy flux  
$\geq 10^{-10} \ \rm erg/cm^2 s$. Therefore   
the data  available in the MeV-GeV band
from binary systems  provide 
important, although not very restrictive  upper limits 
on the energy flux of VHE neutrinos from these objects.
With the arrival of AGILE and especially 
GLAST, the MeV/GeV $\gamma$-ray 
observations will play a more decisive role 
in predictions of VHE neutrinos from compact objects.  

\section{"Orphan" TeV neutrino sources?}

The multiwavelength approach  to the estimates of VHE neutrino fluxes
expected from potential cosmic accelerators indicate that  
the fluxes of the best-candidate \textit {persistent} galactic 
neutrino sources cannot significantly exceed 
$\sim 10^{-11} \ \rm \nu/cm^2 s$, i.e. most likely  
these sources are expected to be revealed at the 
level of statistically marginal signals. 
A possible exception could be "hidden"  sources - 
proton accelerators completely  shielded from us  
by a very thick  ($\gg 100 \ \rm g/cm^2$) gas 
material in which the energy of ultrarelativistic protons 
is converted with 100 \% efficiency to secondaries.
At the same time, in such a thick shell of gas   
the high energy electromagnetic  radiation  
would completely dissipate, and 
thus the source would becomes invisible because of much 
stronger background and forground thermal emission components. 
In such objects, the (highly unknown) 
total acceleration power of  the source is the only model parameter  
that determines  the neutrino flux; the source would be  detectable 
by VHE neutrino detectors if the power of the "hidden" PeVatron exceeds,
approximately by a factor of 10, the estimate given by Eq. (1).

There is another (more sophisticated) scenario  of realization of 
"orphan"  neutrino sources, i.e. objects with  neutrino fluxes not
accompanied by electromagnetic radiation. Such a possibility is related
to the features of acceleration and radiation of particles in  
optically thick (with respect to the photon-photon pair production) 
relativistic flows which can be formed by hot plasma left behind
a relativistic shock or exist in the form of jets.  
Generally,  it is silently  assumed that  nonthermal 
particles and hence their radiation are 
isotropically distributed  in the comoving frame.  
However, this  assumption can easily
be violated in relativistic shocks and jets with a 
strong impact on the emission properties, especially at 
very high energies. Namely, the beam pattern of relativistic jets
with a  bulk motion Lorentz factor $\Gamma$
in this energy domain appears  much broader than
the inverse Lorentz factor, $\Gamma^{-1}$. This results 
in an \textit{off-axis} high energy emission \cite{Der1}  
which is expected to be much brighter compared to the 
predictions derived from the standard Doppler boosting
considerations applied to an isotropic (in the frame 
of the jet)  source.  However, in optically thick 
sources the electromagnetic radiation from super-critical 
particles is reprocessed through the electron-photon 
cascades, thus becomes isotropic 
in the jet frame, and, therefore,   
strongly collimated in the lab-frame. Consequently, the reprocessed
electromagnetic radiation cannot be observable at large viewing angles.
This effect does not concern the high energy neutrinos, therefore   
the jet,  when viewed off-axis, may appear as an \textit{over-bright}
neutrino source with an arbitrarily
large ratio of the neutrino luminosity to
the total electromagnetic luminosity.
It should be noted that acceleration of 
protons in relativistic shocks and shear flows
can be significantly enhanced when it proceeds through the so-called
\textit{converter}  mechanism \cite{Der2}. This acceleration mechanism,  
which utilizes multiple conversions of protons to neutrons 
through photomeson reactions, has certain advantages compared to 
the standard diffusive shock acceleration scenario. It 
essentially  diminishes 
particle losses downstream and provides 
penetration of particles deep into the upstream region  allowing a
highly desirable  energy boost by a factor of $\Gamma^2$ 
at \textit{each} shock encounter \cite{Der2}.  Since the copious neutrino production 
is an intrinsic feature of this scenario,  the realization of 
the convertor acceleration mechanism in relativistic flows 
with large aspect angles   
would naturally lead to the appearance of an "orphan" VHE neutrino source.

\section{First  galactic TeV neutrino sources to be detected...}

A possible  hadronic origin of gamma-radiation of some of the  
TeV $\gamma$-ray  sources  discovered by HESS in the galactic plane  
\cite{FA_Science}
makes them also potential  emitters  of high energy 
neutrinos (see e.g. ref.\cite{Dermer}).  Recently,
Vissani \cite{Vissani} Kistler and Beacom \cite{Beacom} 
and Kappes et al. \cite{Kappes}
performed detailed  calculations of the neutrino 
signal and background rates
for the future 1$\rm km^3$-volume scale  neutrino telescopes based  
on the energy spectra and source-morphologies 
of  galactic TeV $\gamma$-ray sources reported by HESS.  
The potential of the km$^3$-volume 
class detectors is limitted, as mentioned above, by the detection  
of $\geq 1$ TeV neutrino fluxes at the level of  
$\geq 10^{-11} \ \rm \nu/cm^2 s$ confined  within an 
angle $\leq 1^\circ$.  For  power-law 
energy distributions of parent  protons with power-law indices   
$\alpha = 2-3$, the related   $\gamma$-ray flux in the 
same energy interval is slightly, by a factor of 
1 to 2, higher  \cite{Kelner},  i.e. 
quite close to the gamma-ray flux  of the Crab Nebula - 
the standard candle of TeV $\gamma$-ray astronomy.  Thus, the accompanied 
$\gamma$-ray flux  in units of "1 Crab" can be treated  as the detection threshold  of the \textit{galactic} 
neutrino astronomy with km$^3$-volume class instruments.   
Presently, in addition 
to the Crab Nebula itself, three  more TeV $\gamma$-ray 
sources are detected
at the "1 Crab"  level - two young shell-type SNRs RXJ 1713.7-3946 \cite{HESS_RXJ1713}
and RXJ~0852.0-4622 \cite{HESS_VelaJr} (Vela Jr),
as well as  a nearby PWN - Vela X \cite{HESS_VelaX}. 

\subsection{Shell type Supernova Remnants}

Young  Supernova Remnants have been predicted, within a hadronic model,  
as extended TeV $\gamma$-ray and neutrino 
sources with shell type 
morphology and hard energy spectra extending 
to 100 TeV \cite{DAV}. The morphological 
and spectroscopic characteristics of young SNRs 
RXJ~1713.7-3946 and  RXJ~0852.0-4622 reported  
by the HESS collaboration perfectly agree  with these predictions, 
but  ironically cannot yet be considered as  an ultimate proof  the hadronic model.
The "trouble makers"  are  the alternative leptonic models  which relate
the TeV $\gamma$-rays to inverse Compton scattering of electrons responsible 
also for the nonthermal X-ray emission.   Although
these models do not provide satisfactory explanations of 
the energy spectra of $\gamma$-rays, and require very  
low magnetic field of order of 
10 $\mu$G \cite{HESS_RXJ1713,HESS_VelaJr}, 
they unfortunately cannot be safely excluded. 
This prevent us from  
making a robust statement of  detection of  SNRs as sources of cosmic ray protons and nuclei. 
Although it is believed  that detailed theoretical studies of SNRs in the context of their 
multiwavelength properties should allow us to arrive at certain conclusions concerning the 
origin of TeV $\gamma$-ray emission,  
formally only the detection of TeV neutrinos from these 
objects  can be considered  as straightforward and unambiguous proof of acceleration of 
protons and nuclei.  The predicted detection rates of 
$\geq 1$ TeV neutrinos (of order of a few events per one year)
from the  brightest $\gamma$-ray SNRs   RAJ1713.7-3946 
and RAJ0852.0-4622 by  
a  km$^3$-volume detector in the Mediterranean Sea,  make  
the prospects of discovery of TeV neutrinos from these SNRs
rather realistic. Unfortunately the locations of 
these two  SNRs are not favorable for IceCube.  The existence of SNRs of 
similar brightness in TeV $\gamma$-rays located in the Northern Hemisphere 
will be explored,  hopefully soon, by the  VERITAS and MAGIC 
telescope systems.  The 
search for TeV $\gamma$-ray  sources in the Cygnus region - 
one of the most prominent and promising parts  of   the 
galactic plane - is of special interest. The recent Milagro observations of this region 
revealed a diffuse $\gamma$-ray component with several hot spots \cite{Milagro2}, 
the strongest of which, MGRO~J2019+37  could be a neutrino source with a 
flux close to the detection threshold of IceCube \cite{Beacom2}. 

\subsection{Pulsar Wind Nebulae}
 
The close associations  of some of the extended TeV galactic sources discovered by HESS with several  well established synchrotron 
X-ray nebulae  
(MSH~15-52,  PSR~J1826-1334, Vela~X, \textit{etc.}) confirm the early 
theoretical predictions \cite{AhAtKif} on 
TeV gamma-ray visibility of young PWNe 
with spin-down "flux" ($L_0/4 \pi d^2$) exceeding  
$10^{34} \ \rm erg/(kpc^{2} s) $.
The broad-band spectral energy distributions of these sources 
are readily  explained by the standard PWN model  which 
assumes acceleration of ultrarelativistic electrons by the pulsar wind termination shock.
Yet,   in some of these systems  particle acceleration could be
driven by ions present in the relativistic pulsar wind \cite{Arons}. 
These ions are expected to produce $\gamma$-rays and neutrinos via inelastic 
interactons  with the ambient medium 
\cite{Armen,Bednarek2,Bednarek3,Amato}.
In this regard,  the extended TeV source associated with 
the pulsar PSR B0833-45 (Vela~X) is  a possible  candidate for such 
a "hadronic plerion".  Indeed, although the 
observed $\gamma$-ray emission can be 
interpreted as inverse Compton emission of nonthermal electrons \cite{HESS_VelaX},
one needs to make some strong (non-trivial) 
 assumptions in order to explain the rather unusual 
spectrum of this source with photon index $\Gamma \simeq 1.5$ and exponential cutoff around 14 TeV. The steady-state 
electron distribution constrained by $\gamma$-ray data 
requires an $E^{-2}$ type power-law spectrum 
with a sharp cutoff around  70 TeV.
Such a spectrum of electrons can be interpreted 
only in terms of negligible 
synchrotron cooling, which would be possible 
only in the case of unusually low 
nebular magnetic field (a few $\mu$G or less). Moreover, 
the total energy in relativistic electrons and  in 
the magnetic field, which is  required to match the observed 
X-ray and gamma-ray fluxes,
is  only a negligible fraction ($\approx  0.1\%$) 
of the pulsar spin-down energy  
released over the pulsar's life-time $1.1 \times 10^4$ yr. 
This begs the question  as to  
where  the remaining energy has gone?  
Interestingly if we assume that a large  fraction 
of the spin-down luminosity of the pulsar is carried 
out by  relativistic protons and nuclei,
one can satisfactorily explain 
both the absolute flux and the spectrum of TeV $\gamma$-rays of 
this unusual source \cite{Horns}. 
Remarkably, the TeV neutrino flux  expected within this scenario should be  
detectable by KM3NeT \cite{Kappes}. This makes the Vela X as one of 
the best-bet candidates  to be the first \textit{detected} astronomical 
TeV neutrino source.  For the IceCube detector an obvious  target  representing 
this source population  is the Crab Nebula.

\subsection{Compact Binary Systems}

The recent detections  of TeV $\gamma$-rays from two binary systems tentatively called 
microquasars -- LS 5039 by HESS \cite{HESS_LS} and LSI 61 303 by MAGIC \cite{MAGIC_LS} --
are amongst the most exciting discoveries of observational gamma-ray astronomy in the very high energy regime. This result clearly demonstrates that the galactic binary systems 
containing a luminous optical star and a compact object
(a black hole or a pulsar/neutron star), 
are sites of effective acceleration of particles (electrons and/or protons) to multi-TeV energies. 
As usual, whether the $\gamma$-rays are of hadronic or leptonic origin is a key question
which however does not have a  straightforward answer (see e.g. ref. \cite{Paredes}).  
The critical analysis  of conditions of  particle acceleration and radiation 
in these sources, based on the temporal and spectral behavior
of TeV $\gamma$-ray emission, in particular on the
modulation of the TeV flux of LS 5039 
with a period of 3.9 day, and the extension of its energy spectrum to 10 TeV and beyond, reduces the possible interpretations  
to a few options. One of them gives a 
preference to the  hadronic origin of TeV photons, 
especially if they are produced within the binary system~\cite{Mitya}. 
If so, the detected $\gamma$-rays should be accompanied by a flux of high energy neutrinos emerging from the decays of $\pi^\pm$ mesons produced 
by \textit{p-p} and/or $p \gamma$ interactions. 
The neutrino fluxes, which can be estimated on the basis of the detected TeV $\gamma$-ray fluxes, taking into account the severe internal $\gamma \gamma \to e^+e^-$ absorption, depend significantly on the location of $\gamma$-ray production region(s) \cite{Dubus,Markus}. 
The minimum neutrino flux above 1 TeV is expected to be at the level of $10^{-12}  \rm \ \nu/cm^{2} s$; however, it could be 
much higher  -  by a factor of 10, or even more. The detectability of 
the TeV neutrino signals significantly
depends on the high energy cutoff in the spectrum of parent protons.
If the spectrum of accelerated protons continues to 100 TeV and 
beyond, the predicted neutrino fluxes of 
LS~5039  and LSI~61~303 
can be probed  by  KM3NeT \cite{Mitya} and 
IceCube  \cite{Gustavo,Diego} high energy neutrino detectors.

\end{document}